\documentclass[twocolumn,showpacs,preprintnumbers,superscriptaddress,amsmath,amssymb]{revtex4}
\usepackage{graphicx}
\usepackage{dcolumn}
\usepackage{bm}

\begin{document}

\title{Magnetowave Induced Plasma Wakefield Acceleration for \\
Ultra High Energy Cosmic Rays}
\author{Feng-Yin Chang}
\affiliation{Kavli Institute for Particle Astrophysics and
Cosmology, Stanford Linear Accelerator Center, Stanford
University, Stanford, CA 94305, USA.}\affiliation{Institute of
Physics, National Chiao-Tung University, Hsinchu 300, Taiwan.}
\author{Pisin Chen}%
\email{chen@slac.stanford.edu} \affiliation{Kavli Institute for
Particle Astrophysics and Cosmology, Stanford Linear Accelerator
Center, Stanford University, Stanford, CA 94305,
USA.}\affiliation{Institute for Astrophysics, National Taiwan
University, Taipei 106, Taiwan.}
\author{Guey-Lin Lin}
\affiliation{Institute of Physics, National Chiao-Tung University,
Hsinchu 300, Taiwan.}
\author{Kevin Reil}
\affiliation{Kavli Institute for Particle Astrophysics and
Cosmology, Stanford Linear Accelerator Center, Stanford
University, Stanford, CA 94305, USA.}
\author{Richard Sydora}
\affiliation{Department of Physics, University of Alberta,
Edmonton, Alberta, Canada.}


\maketitle

{\bf Magnetowave induced plasma wakefield acceleration (MPWA) in a
relativistic astrophysical outflow has been proposed as a viable
mechanism for the acceleration of cosmic particles to ultra high
energies\cite{plasma:Chen02}. Here we present simulation results
that clearly demonstrate the viability of this mechanism for the
first time. We invoke the high frequency and high speed whistler
mode for the driving pulse. The plasma wakefield so induced
validates precisely the theoretical prediction. We show that under
appropriate conditions, the plasma wakefield maintains very high
coherence and can sustain high-gradient acceleration over a
macroscopic distance. Invoking gamma ray burst (GRB) as the
source, we show that MPWA
production of ultra high energy cosmic rays (UHECR) beyond ZeV $(10^{21}{\rm eV})$ is possible.}\\

The origin of ultra high energy cosmic rays (UHECR) is a long-standing mystery in astrophysics. Thus far, the theories that attempt to explain the origin of UHECR can be broadly categorized into the ``top-down" and the ``bottom-up" scenarios.
Each scenario faces its own theoretical and observational challenges\cite{Olinto}. Precision measurements\cite{FLASH,AIRFLY} on the yield of air-shower induced fluorescence lend support to the energy calibration of the HiRes observations\cite{cosmic:HiRes}. Recent data from the Auger Observatory\cite{cosmic:Auger}, also fluorescence normalized, exhibit a similar location of the ``ankle" and super-GZK steepening as HiRes. The lack of a strong super-GZK flux reduces the need for top-down exotic models. If these models are indeed disfavored, then the challenge to find a viable ``bottom up" mechanism to accelerate ordinary particles to and beyond $10^{20}$ eV becomes more acute.

Shocks, unipolar inductors and magnetic flares are the three most
potent, observed, ``conventional" accelerators that can be
extended to account for $\sim{\rm ZeV} (=10^{21}{\rm eV})$ energy
cosmic rays\cite{Blandford:1999}. Radio jet termination shocks and
gamma ray bursts (GRB) have been invoked as sites for the shock
acceleration, while dormant galactic center black holes and
magnetars have been proposed as sites for the unipolar inductor
acceleration and the flare acceleration, respectively. Each of
these models, however, presents problems\cite{Blandford:1999}.
Evidently, novel acceleration mechanisms that can avoid the
difficulties faced by these conventional models should not be
overlooked.

Plasma wakefield accelerators\cite{LWFA:Tajima,PWFA:Chen85} are
known to possess two salient features: (1) The energy gain per
unit distance does not depend (inversely) on the particle's
instantaneous energy or momentum. This is essential to avoid the
gradual decrease of efficiency in reaching ultra high energies;
(2) The acceleration is linear. Bending of the trajectory is not a
necessary condition for this mechanism. This helps to minimize
inherent energy loss which would be severe at ultra high energy.
However, high-intensity, ultra-short photon or particle beam
pulses that excite the laboratory plasma wakefields are not
readily available in the astrophysical setting. It was, however,
proposed\cite{plasma:Chen02} that large amplitude plasma
wakefields can instead be excited by the astrophysically more
abundant plasma ``magnetowaves", whose field components are
magnetic in nature ($|B|>|E|$). Protons can be accelerated beyond
ZeV energy by riding on such wakefields. Though attractive, this
concept has never been validated through self-consistent computer
simulations. In this Letter, we report on the plasma
particle-in-cell (PIC) simulation results that confirm the
magnetowave-induced plasma wakefield acceleration (MPWA) concept
for the first time.

Magnetized plasmas support a variety of wave modes propagating at
arbitrary angles to the imposed magnetic field. For the specific
case of wave modes propagating parallel to the external magnetic
field, the electromagnetic waves become circularly polarized and
the dispersion relation is \cite{Stix}
\begin{equation}
 \omega^{2}=k^{2}c^{2}+\frac{\omega_{ip}^{2}}{1\pm\omega_{ic}/\omega}+
\frac{\omega_{p}^{2}}{1\mp\omega_{c}/\omega}\, ,\label{dispersion}
\end{equation}
where the upper (lower) signs denote the right-hand (left-hand)
circularly polarized waves. $\omega_{p}=\sqrt{4\pi e^2 n_{p}/m_e}$
is the electron plasma frequency, $\omega_{c} = eB/m_ec$ is the
electron cyclotron frequency and the subscript $i$ denotes the ion
species. Each polarization has two real solutions with high and
low frequency branches and both have a frequency cutoff which
forms a forbidden gap for wave propagation. The right-hand
polarized, low frequency solution is called the whistler wave
which propagates at a phase velocity less than the speed of light.
When the magnetic field is sufficiently strong such that
$\omega_{c}\gg \omega_{p}$, the dispersion of the whistler mode
becomes more linear over a wider range of wavenumbers with phase
velocity approaching the speed of light (see Fig.1). The $E$ and
$B$ components of the wave are now comparable in strength. In this
regime the travelling wave pulses can maintain their shape over
macroscopic distance, a condition desirable for plasma wakefield
acceleration.

The ponderomotive force in a magnetized plasma has been well
studied\cite{wk}. Applying the dispersion relation for the
whistler wave, with the ion motion neglected, we obtain the
ponderomotive force acting on an individual electron as
\begin{equation}
F_{z}=-\frac{1}{2}\frac{e^2}{m_e\omega
(\omega-\omega_{c})}\left[1+\frac{kv_g\omega_c}{\omega(\omega-\omega_c)}\right]\partial_{\zeta}E_{_W}^2(\zeta),
\end{equation}
where $E_{_W}(\zeta)$ is the amplitude of the whistler
wave-packet, and $\zeta\equiv z-v_gt$ the co-moving coordinate for
the driving pulse. Note that $E_{_W}$ is perpendicular to $z$.

Combining this equation with the continuity equation and the
Poisson equation, the longitudinal electric field in the plasma,
i.e., the plasma wakefield, can be solved and it reads
\begin{eqnarray}
E_z(\zeta)=-\frac{ek_pE_{_W}^2}
{m_e\omega(\omega-\omega_c)}\left[1+\frac{kv_g\omega_c}{\omega(\omega-\omega_c)}\right]
\chi(\zeta), \label{wakefield}
\end{eqnarray}
with
\begin{equation}
\chi(\zeta)=\frac{k_p}{2E_{_W}^2}\int_{\zeta}^{\infty}d\zeta^{\prime}E_{_W}^2(\zeta^{\prime})
\cos\left[k_p(\zeta-\zeta^{\prime})\right],
\end{equation}
where $E_{_W}$ is the maximum value of $E_{_W}(\zeta)$. An
expression similar to Eq.~(\ref{wakefield}) has been obtained for
laser-induced wakefield in a magnetized plasma\cite{sh}. For a
Gaussian driving pulse with $E_{_W}(\zeta)=E_{_W}
\exp(-\zeta^2/2\sigma^2)$, it can be shown that behind the driving
pulse, i. e., $|\zeta| \gg \sigma$,
\begin{eqnarray}
\chi(\zeta)=\frac{\sqrt{\pi}}{2} k_p\sigma
e^{-k_p^2\sigma^2/4}\cos k_p\zeta \equiv \chi\cos k_p\zeta.
\end{eqnarray}
It is customary to express the plasma wakefield in terms of the
Lorentz invariant ``strength parameter" of the driving pulse,
$a_0\equiv eE_{_W}/m_ec\omega$, and the ``wavebreaking" field,
$E_{wb}\equiv m_ec\omega_p/e$. Assuming the driving pulse
frequency is centered around $\omega$ and its speed $v_{g}\approx
\omega/k$, the maximum wakefield, or the {\it acceleration
gradient}, attainable behind the driving pulse is then
\begin{equation}
G=\chi\frac{k^{2}c^2}{(\omega-\omega_{c})^{2}} a_0^{2}E_{wb}\,
,\quad a_0\ll 1\, . \label{wakefield3}
\end{equation}
Relative to the conventional wakefields, $G$ is enhanced by a
factor $k^2c^2/(\omega-\omega_c)^2$ when $\omega$ approaches
$\omega_c$. The above expression is derived under the assumption
of linear plasma perturbation, i.e., $a_0 \ll 1$. Research made
over the past two decades in plasma wakefields has firmly
established the generalized wakefield amplitude for all values of
$a_0$\cite{plasma:Esarey}, and adapting its form we obtain,
\begin{equation}
G=\chi
\frac{k^{2}c^2}{(\omega-\omega_{c})^{2}}\frac{a_0^2}{\sqrt{1+a_0^2}}E_{wb}\,.
\label{wakefield2}
\end{equation}
Note that in the nonlinear regime ($a_0\gg 1$) the wakefield is
no longer sinusoidal in $\zeta$ but saw-tooth-like.

We have conducted computer simulations to study the MPWA process
driven by a Gaussian driving whistler pulse described above. Our
simulation model integrates the relativistic Newton-Lorentz
equations of motion in the self-consistent electric and magnetic
fields determined by the solution to Maxwell's
equations\cite{Dawson, Sydora}. The 4-dimensional phase space
$(z,p_x,p_y,p_z)$ is used for the charged particle dynamics and a
uniform external magnetic field, $B_0$, is imposed in the
z-direction. In order to sustain the driving pulse shape
propagating in the plasma, it is important to have modes of the
pulse travelling with similar phase velocities. Therefore, we used
a wavepacket with Gaussian width $\sigma =80\Delta/\sqrt{2}$,
where $\Delta$ is the cell size taken to be unity, and the
wavenumber $k = 2\pi/60\Delta$. The normalized physical parameters
$\omega_{c}/ \omega_{p}=6$ and $m_i/m_e=2000$ were taken and for a
uniform background plasma with electron collisionless skin depth,
$c/\omega_{p}\Delta=30$, this gives $\omega/\omega_{p}=2.98$ and
$v_{g}/c\simeq \omega/ck= 0.95$. Other numerical parameters used
are: total number of cells in the $z$-direction, $L_z
=8192\Delta=273c/\omega_p$, average number of particles per cell
was 10, and the time step $\omega_p\Delta t=0.1$. The fields were
normalized by $(1/30)E_{wb}$.

\begin{figure}[tb]
\begin{center}
$\begin{array}{c}
\includegraphics[width=8.5cm]{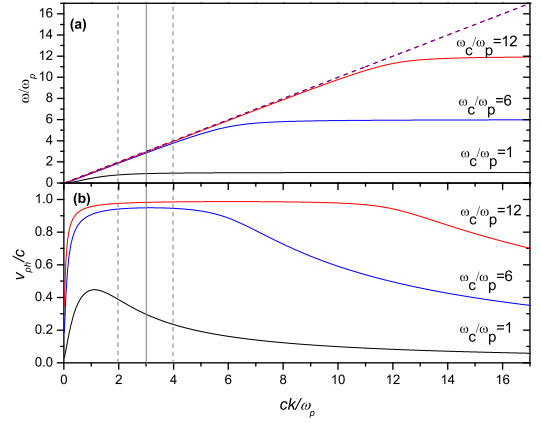}
\end{array}$
\end{center}
\caption{(a) Frequency and (b) phase velocity versus wavenumber
for different magnetic field strengths. The vertical solid line is
the mean value of the pulse wavenumber that was chosen for the PIC
simulation, and the dashed lines its range.} \label{pulse}
\end{figure}

We set the maximum amplitude $E_{_W}=10$, which gives the
normalized vector potential $a_{0}=eE_{_W}/m_ec\omega=0.11\ll 1$.
Thus the wakefield in our simulation is in the linear regime. The
pulse was initialized at $z_0=500\Delta=16.66 c/\omega_p$. To
avoid spurious effects, we gradually ramped up the driving pulse
amplitude until $t=100 \omega_{p}^{-1}$, during which the plasma
feedback to the driving pulse was ignored. After this time, the
driving pulse-plasma interaction was tracked self-consistently. As
the dispersion relation in this regime is not perfectly linear,
there was a gradual spread of the pulse width. Thus $\chi$ and
$E_{_W}$ of the driving pulse decrease accordingly. As a result,
the maximum wakefield amplitude, $E_z$, declined in time. Even so,
it agrees very well with the theoretical maximum of $E_{z}\sim
0.266(1/30)E_{wb}$. Fig.2 is a snapshot of $E_{x}$ and $E_{z}$ at
$t=230\omega_{p}^{-1}$. We note that while the driving pulse
continues to disperse, the wakefield remains extremely coherent.

\begin{figure}[tb]
\begin{center}
$\begin{array}{c}
\includegraphics[width=8.5cm]{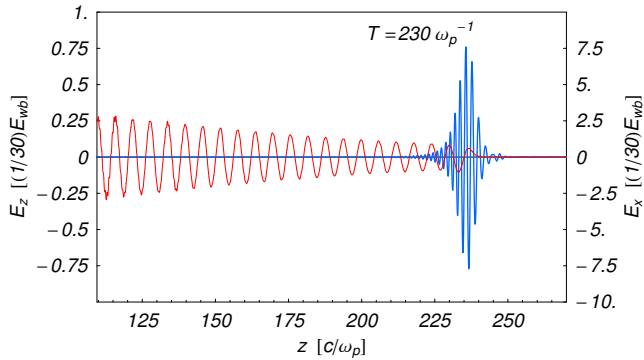}
\end{array}$
\end{center}
\caption{A snapshot of the plasma wakefield induced by the
whistler pulse. $E_x$ is in blue and $E_z$ in red.}
\label{wakeplot}
\end{figure}

\begin{figure}[tb]
\begin{center}
$\begin{array}{c}
\includegraphics[width=8.5cm]{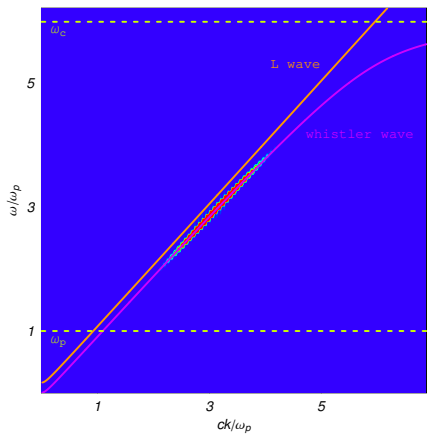}
\end{array}$
\end{center}
\caption{The dispersion relation of the driving pulse from PIC
simulation. Liner dispersion curves for the L-wave (orange) and
the whistler wave (pink) are superimposed.} \label{simdisp}
\end{figure}

We sampled the $E_{x}(k)$ of the pulse, after its initialization,
at every time step and analyzed it in the frequency space.
Fig.\ref{simdisp} shows the $\omega -k$ intensity generated from
the PIC simulation driving pulse power spectrum. It is
superimposed with the theoretical curves for the left-handed
circularly polarized electromagnetic wave (L-wave) and the
whistler wave dispersion relations. We confirm that our driving
pulse is indeed a whistler wave.

\begin{figure}[tb]
\begin{center}
$\begin{array}{c}
\includegraphics[width=8.5cm]{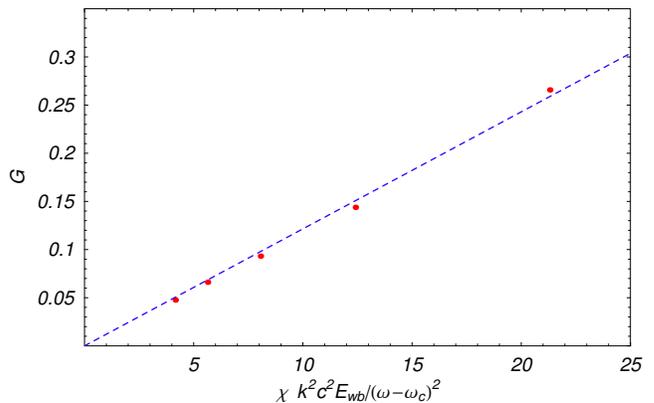}
\end{array}$
\end{center}
\caption{Validation of the functional dependence of $G$ in
Eq.(\ref{wakefield3}), where $E_{wb}$ is in units of
$(1/30)m_ec\omega_p/e$.} \label{bEmax}
\end{figure}

Next we validate the functional dependence of the acceleration
gradient given in Eq.(\ref{wakefield3}). Fig.\ref{bEmax} plots the
acceleration gradient versus $\chi
k^2c^2E_{wb}/(\omega-\omega_{c})^{2}$. This is performed by
varying $\omega_c$. Wavepackets with the same wavenumber
($2\pi/60\Delta$) and the same maximum amplitude ($E_{_W}=10$) are
initialized with different $B_0$ such that
$\omega_{c}/\omega_{p}=6, 7, 8, 9, 10$. The linear fitting slope
is 0.012, which agrees very well with the expected value of
$a_0^2$. We comment that when $\omega$ approaches $\omega_{c}$, i.
e., when the two are in resonance, the wakefield amplitude can be
dramatically increased by several orders of magnitude. In view of
the successful validation by our computer simulation of the MPWA
in the linear regime, we are safe to extrapolate it to the
nonlinear regime according to the theoretical formula in
Eq.(\ref{wakefield2}).

We now relate this mechanism to the issue of UHECR acceleration.
An earlier attempt has been made\cite{plasma:Chen02} where gamma
ray bursts (GRB) were invoked as the candidate site for plasma
wakefield production of UHECR. We follow the same approach here.
GRBs are generally classified into long ($\tau_l\sim 10-100$sec)
and short ($\tau_s\sim 1$sec) bursts. Within seconds (for short
bursts), about $\mathcal{E}_{\rm GRB}\sim 10^{50}$erg of energy is
released through gamma rays. We invoke the Neutron Star-Neutron
Star coalescence as our working model for GRB. Neutron stars are
known to be compact ($R_{\rm NS}\sim \mathcal{O}(10)$km) and carry
intense surface magnetic fields ($B_{\rm NS}\sim 10^{13}$G). It is
generally believed that when neutron stars collide, the tremendous
release of energy creates a highly relativistic out-bursting
fireball (jets)\cite{GRB:Mezaros}, most likely in the form of a
plasma. We assume such a jet has an open-angle of $\theta_{\rm
GRB}\sim 0.1$ and the initial plasma density in the jet $n_{\rm
GRB}\sim 10^{26}{\rm cm}^{-3}$. We further assume that such
violent collision of intense magnetic fields would create sequence
of strong magnetoshocks, where whistler waves are imbedded. In the
aftermath of such tremendous impact, the magnetic field-lines
would be temporarily shattered and reoriented with strong poloidal
field lines parallel to the axis of the jet.

From the previous discussion we see that the MPWA is most
effective when the driving pulse frequency falls between the
plasma frequency, $\omega_p$, and the electron-cyclotron
frequency, $\omega_c$. Let us verify whether and where this
condition can be satisfied along the GRB jet. First we note that
due to the conservation of the magnetic flux, the poloidal
magnetic field strength decreases as $1/r^2$ away from the
epicenter of GRB. This means $\omega_c=eB/m_ec\propto (R_{\rm
NS}/r)^2$. On the other hand, the continuity condition requires
that the plasma density decreases as $1/r^2$ as well. Therefore,
$\omega_p\propto R_{\rm NS}/r$ and the cross-over between these
two parameters does exist at a distance $R\sim 100R_{\rm NS}\sim
1000{\rm km}$. This is the ``sweet spot" where MPWA is the most
effective.

To estimate the plasma wakefield acceleration gradient at this
``sweet spot", we first note the EM energy density of GRB is
$E_{\rm GRB}^2/4\pi=\mathcal{E}_{\rm GRB}/c\tau_s\pi(\theta_{\rm
GRB}R)^2\sim 10^{27}{\rm erg/cm}^3$. We assume that a fraction,
$\eta_a$, of this outburst energy goes into the magnetoshocks. We
further assume that a fraction, $\eta_b$, of the magnetoshocks
energy lies in the whistler mode. This means $E_{_W}^2\sim
\eta_a\eta_b E_{\rm GRB}^2$. Based on this, the associated
strength parameter $a_0$ is: $a_0=\sqrt{\eta_a\eta_b}(eE_{\rm
GRB})/(m_ec\omega).$ At the sweet spot where
$\omega_c\sim\omega\simeq kc\sim\omega_p$, the factor
$k^2c^2/(\omega^2-\omega_{c}^2)$ is of the order unity.
Furthermore, the extremely sharp magnetoshock fronts would render
the form factor $\chi$ also of the order unity. Assume that $a_0
>1$. Then the acceleration gradient boils down to, {\it cf.}
Eq.(\ref{wakefield2}),
\begin{equation}
G\sim a_0E_{wb}=\sqrt{\eta_a\eta_b}\Big(\frac{eE_{\rm
GRB}}{m_ec\omega}\Big)E_{wb}\, .
\end{equation}
To appreciate what this translates into physical requirements, let
us assume that the range of the ``sweet spot" is $\delta R\sim
0.1R\sim 10R_{\rm NS}\sim 100{\rm km}$ around $R$ where the factor
$k^2c^2/(\omega^2-\omega_{c}^2)$ is of the order unity. Then in
order for MPWA to be responsible for the production of UHECR
beyond ZeV ($10^{21}$eV), it is necessary that $G\sim 10^{14}{\rm
eV/cm}$. In turn, the fractions of GRB energy imparted into the
whistler mode have to be $\eta_a \sim \eta_b\sim 10^{-2}$.

As shown in Ref.\cite{plasma:Chen02}, the stochastic encounters of
the test particle with the random acceleration-deceleration phases
would result in a inverse-square-law spectrum,
$f(\mathcal{E})\propto 1/\mathcal{E}^{2}$. The various additional
energy loss mechanisms, such as few-body collision and synchrotron
radiation, would degrade the power-law index to
$1/\mathcal{E}^{2+\beta}$, with $0 <\beta < 1$.

Our PIC simulations have confirmed the concept of plasma wakefield
excited by a magnetowave in a magnetized plasma. Different from
the laser and particle beam, the magnetowaves are medium waves
which cannot exist without the plasma. MPWA should thus be of
interest as a fundamental phenomena in plasma physics and an
alternative approach to plasma wakefield acceleration.

As a first step, we investigated MPWA in the parallel-field
configuration. Since both poloidal and toroidal field components
are inevitable in astro-jets, we will further investigate plasma
wakefield excitation and acceleration under the cross-field
configuration. In order for MPWA to be responsible for the ZeV
UHECR production, the energy transfer efficiency for GRB is
constrained. It would be very interesting both observationally and
theoretically to test whether this constraint is valid. \\

\noindent{\bf Acknowledgements} We thank R. Noble for valuable
suggestions. GLL appreciates the hospitality of Kavli Institute
for Particle Astrophysics and Cosmology at SLAC. This work is
supported by US DOE (Contract No. DE-AC03-76SF00515), National
Science Council of Taiwan (Grant No. 95-2119-M-009-026), and
Natural Sciences and Engineering Research Council of Canada.

\end{document}